\documentclass[fleqn,twoside]{article}
\usepackage{espcrc2}
\usepackage{graphicx}

\def\dmatm{$\Delta m^2_{32}$}
\def\dmsol{$\Delta m^2_{21}$}
\def\sstt23{$\sin^2 2\theta_{23}$}
\def\numutonue{$\nu_\mu \to \nu_e$}

\def\cerenkov{Cherenkov}

\newlength{\overlaywd}
\newlength{\overlayht}
\newcommand{\overlay}[4]{%
  \settowidth{\overlaywd}{#2}\settoheight{\overlayht}{#2}%
  \raisebox{#4\overlayht}{%
    \makebox[0pt][l]{\hspace{#3\overlaywd}%
      #1}}%
  #2}%

\title{Very Long Baseline Neutrino Oscillations, The BNL VLBNO
  Concept.}
\author{Brett Viren\address{Brookhaven National Laboratory,\\
    510E Physics, Upton NY, USA, 11973-5000}}
\begin{document}

\begin{abstract}
  A wide energy-band neutrino beam sent over a very long baseline to a
  massive detector can break the degeneracies in the neutrino
  oscillation parameters.  It can measure the disappearance parameters
  with precision and determine the mass hierarchy.  If $\theta_{13}$
  is large enough the CP violating phase can be measured with neutrino
  running alone and anti-neutrino running can confirm CPV and improve
  the parameter measurements.  Brookhaven National Laboratory is
  pursuing such an experiment.
 \vspace{1pc}
\end{abstract}
\maketitle

\section{Introduction}

As shown at this conference and elsewhere, one of the main
challenges in measuring neutrino oscillation parameters in most
current and proposed experiments are degeneracies.

Brookhaven National Laboratory is designing an experiment that can
break these degeneracies, measure or limit all neutrino parameters,
determine the mass hierarchy and be sensitive to new physics.  This
experiment incorporates intense, wide band and high energy neutrino
and anti-neutrino beams, a very long baseline and a massive far
detector.

\section{Motivation for Experiment Parameters}

The primary motivation for this arrangement is to observe multiple
neutrino oscillation periods.  This provides two main benefits over
traditional single oscillation LBNO experiments.  First, since
peak-to-valley and node-to-node features of the disappearance
oscillation pattern can be resolved, a precision measurement of
\dmatm{} and \sstt23{} can be made which is not strongly dependent on
absolute event normalization.  Second, the effects that govern
\numutonue{} appearance have different strengths at different energies
so by resolving multiple appearance peaks these, otherwise degenerate
effects, can be disentangled.

Fermi motion of the nuclei in the active medium of the far detector
begins to dominate the reconstructed energy resolution when the
neutrino energy is below $\sim 500$ MeV.  Avoiding this requires high
neutrino energies and placing the far detector at a baseline that is
long enough for multiple oscillation periods to occur well away from
this Fermi motion dominated energy region.  Current best fits of
\dmatm implies a baseline of $>2000$ km and an energy coverage from
above the Fermi-motion dominated range up to $\sim$10 GeV.

Finally, in order to collect enough statistics a massive detector, a
high intensity neutrino source and a sufficiently long running time
are needed.  We assume a Water \cerenkov{} detector of 500 kTon
fiducial mass, such as UNO or HyperK with performance as good or
slightly better than current Super-Kamiokande.

An upgrade\cite{agsupgrade} to the BNL Alternating Gradient
Synchrotron (AGS) will initially produce a 1 MW proton beam.  
The Booster will be replaced by a 1.2 GeV super-conducting Linac that
will increase the protons per pulse from $7$ to $9\times10^{13}$ and
allow the fill time to be reduced from 0.6 second to 1.0 millisecond.
Power supply and magnet upgrades will be needed to improve the AGS
repetition rate from 0.5 Hz to 2.5 Hz.

\begin{figure}[htb]
\begin{center}
  \includegraphics*[width=0.4\textwidth]{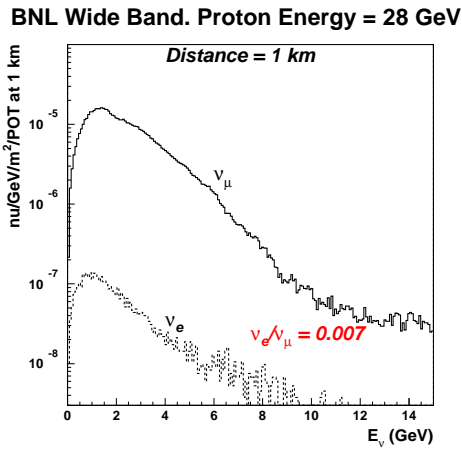}
\end{center}
\vspace{-1cm}
\caption{Expected neutrino flux at 1 km with 1 MW, 28 GeV proton beam, 60 cm carbon target and 4 m wide, 200 m long decay tunnel.}
\label{ags}
\end{figure}

The proton beam will be directed to a fixed target and conventional
focusing horn system positioned on a $\sim$50 m tall hill.  The
secondaries will decay down a 200 m long, 4 m wide tunnel pointing
towards the far detector producing the expected neutrino flux shown in
Fig. \ref{ags}.  The far site is assumed to be either of the two
DUSEL candidates, Homestake, SD or Henderson, CO at 2540 km and 2770
km from BNL, respectively.  Initial running will determine the mass
hierarchy and dictate the subsequent running mode.  The nominal run
plan is 5 years\footnote{We take 1 year = $10^7$ seconds.}  neutrino
running at 1 MW followed by 5 years anti-neutrino running at 2 MW.

\section{Illustrative Oscillation Probability Plots}

This presentation attempts to focus on how the degeneracies in the
parameters governing \numutonue{} appearance are broken through
examining simple oscillation probability plots.  The cost of this
simplicity is to ignore the extremely critical issues of detector
performance.  These issues are being addressed in the larger context
of the concept and other work\cite{milind,chiaki} has gone in to more
detail.

Unless otherwise stated, for these probabilities plots the nominal
values are taken to be there current best fits:
$\Delta m^2_{21}$ = 8.0e-5 $e$V$^2$,
$\Delta m^2_{32}$ = 2.5e-3 $e$V$^2$,
$\sin^2(2\theta_{23})$ = 1.0,
$\sin^2(2\theta_{12})$ = 0.86. 
For the unknown values we take
$\sin^2(2\theta_{13})$ = 0.04 
and
$\delta_{CP}$ = 0.
The baseline is taken to be the BNL-Homestake one of 2540 km.  The
longer BNL-Henderson is essentially equivalent.  The neutrinos are
propagated through the PREM\cite{prem} Earth density profile.

\begin{figure}[htb]
\begin{minipage}{0.45\textwidth}
\overlay{{\tiny (a)}}{\includegraphics*[width=0.5\textwidth]{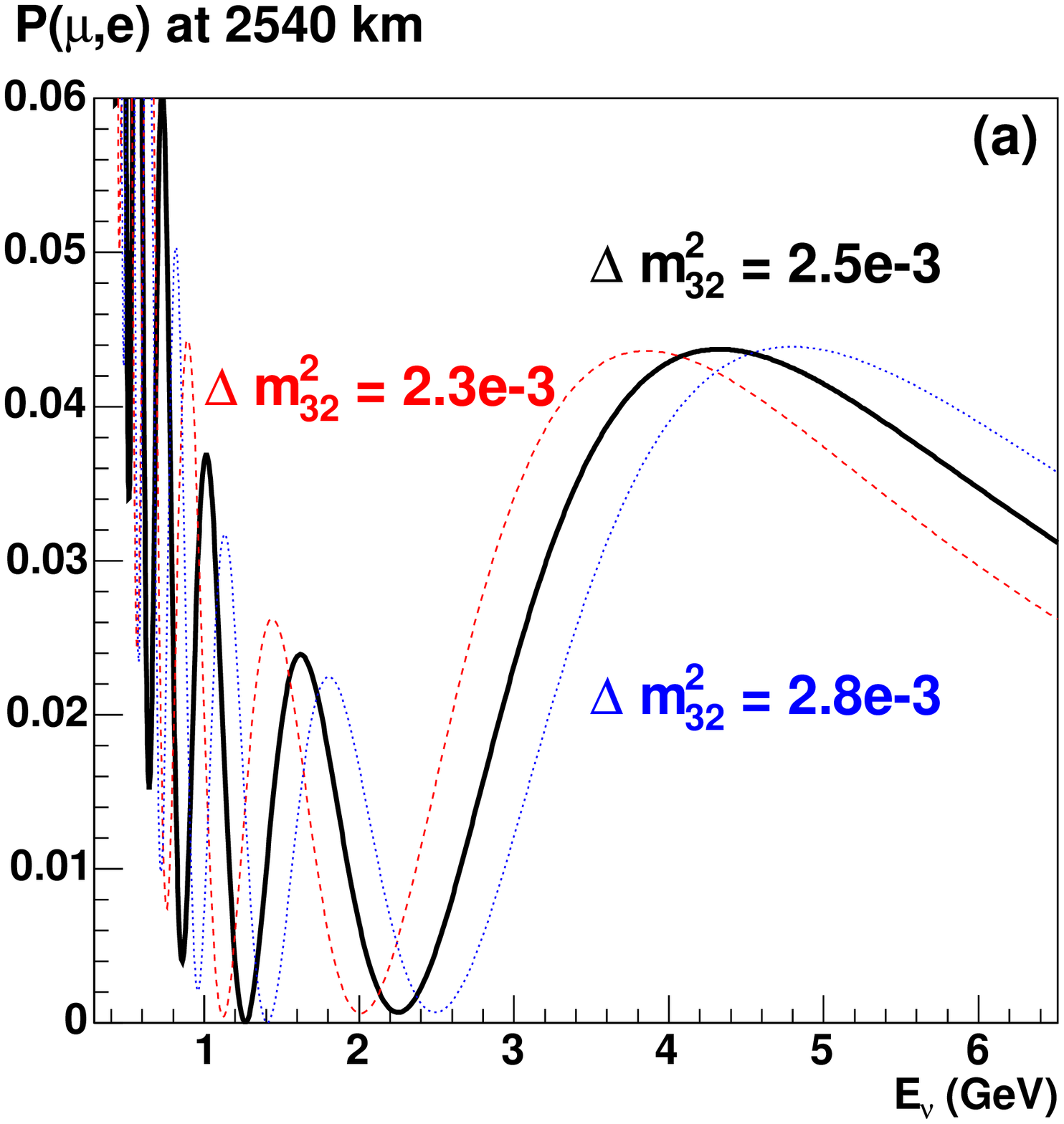}}{0.8}{0.8}%
\overlay{{\tiny (b)}}{\includegraphics*[width=0.5\textwidth]{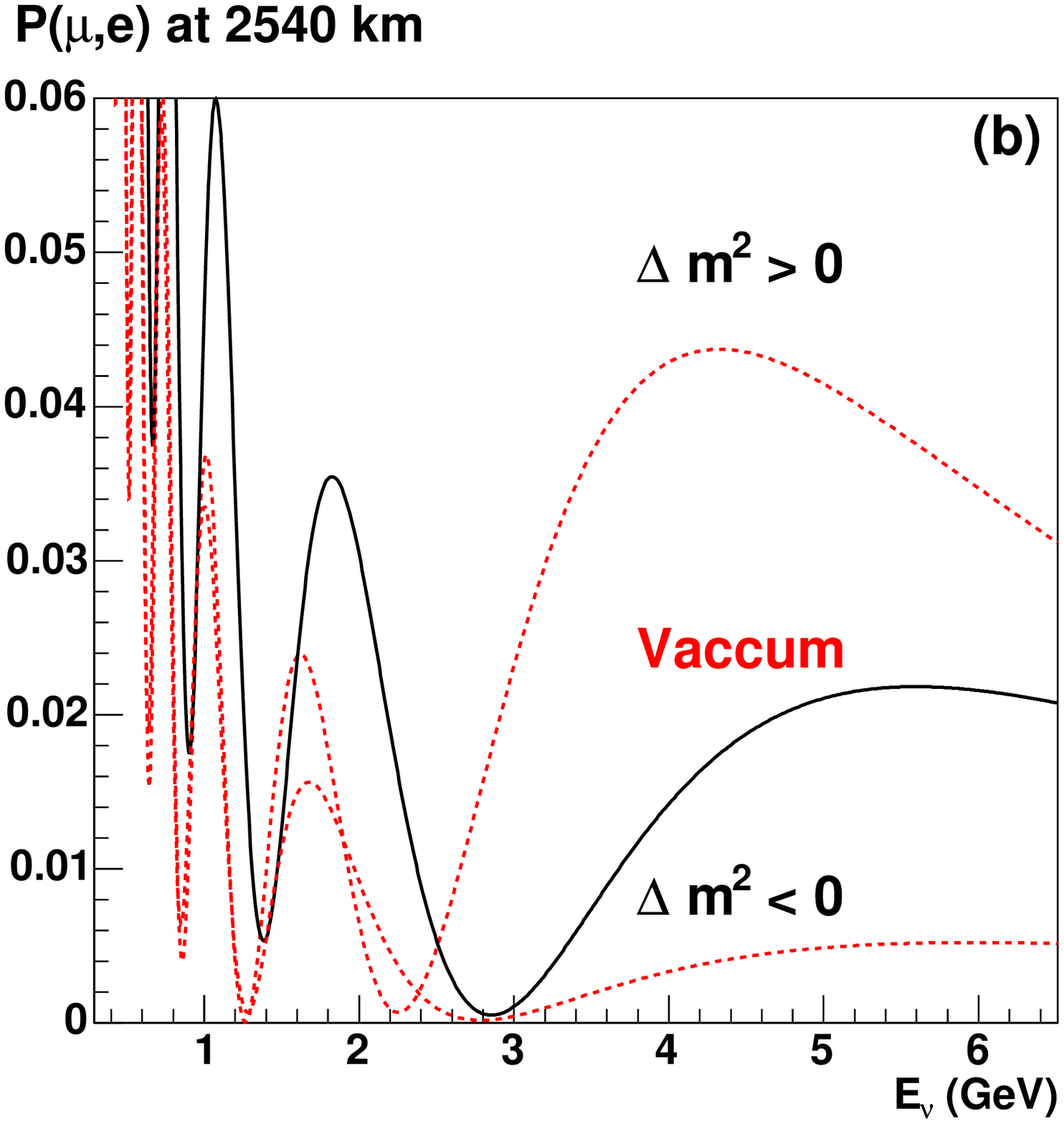}}{0.8}{0.8}

\overlay{{\tiny (c)}}{\includegraphics*[width=0.5\textwidth]{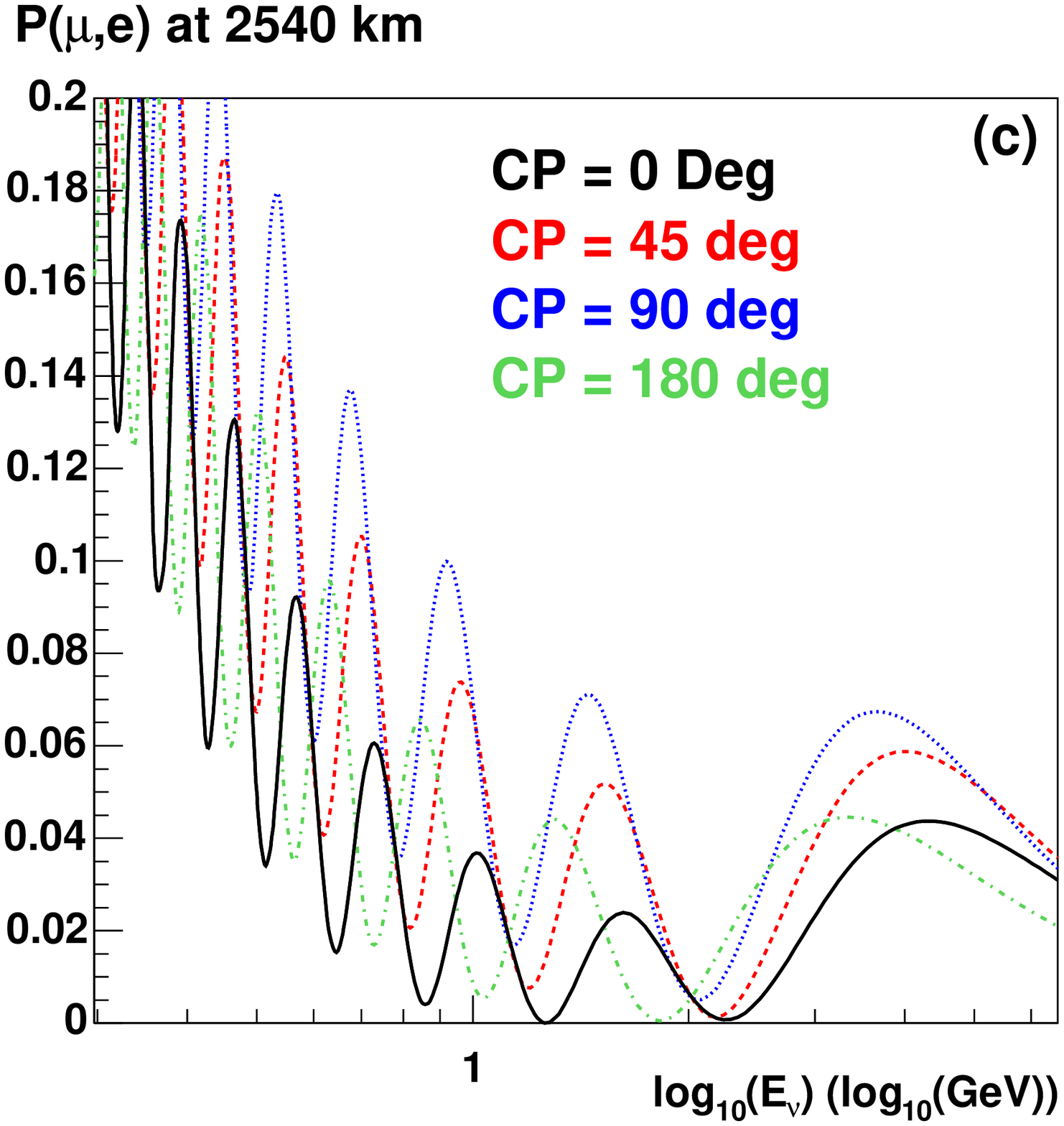}}{0.8}{0.8}%
\overlay{{\tiny (d)}}{\includegraphics*[width=0.5\textwidth]{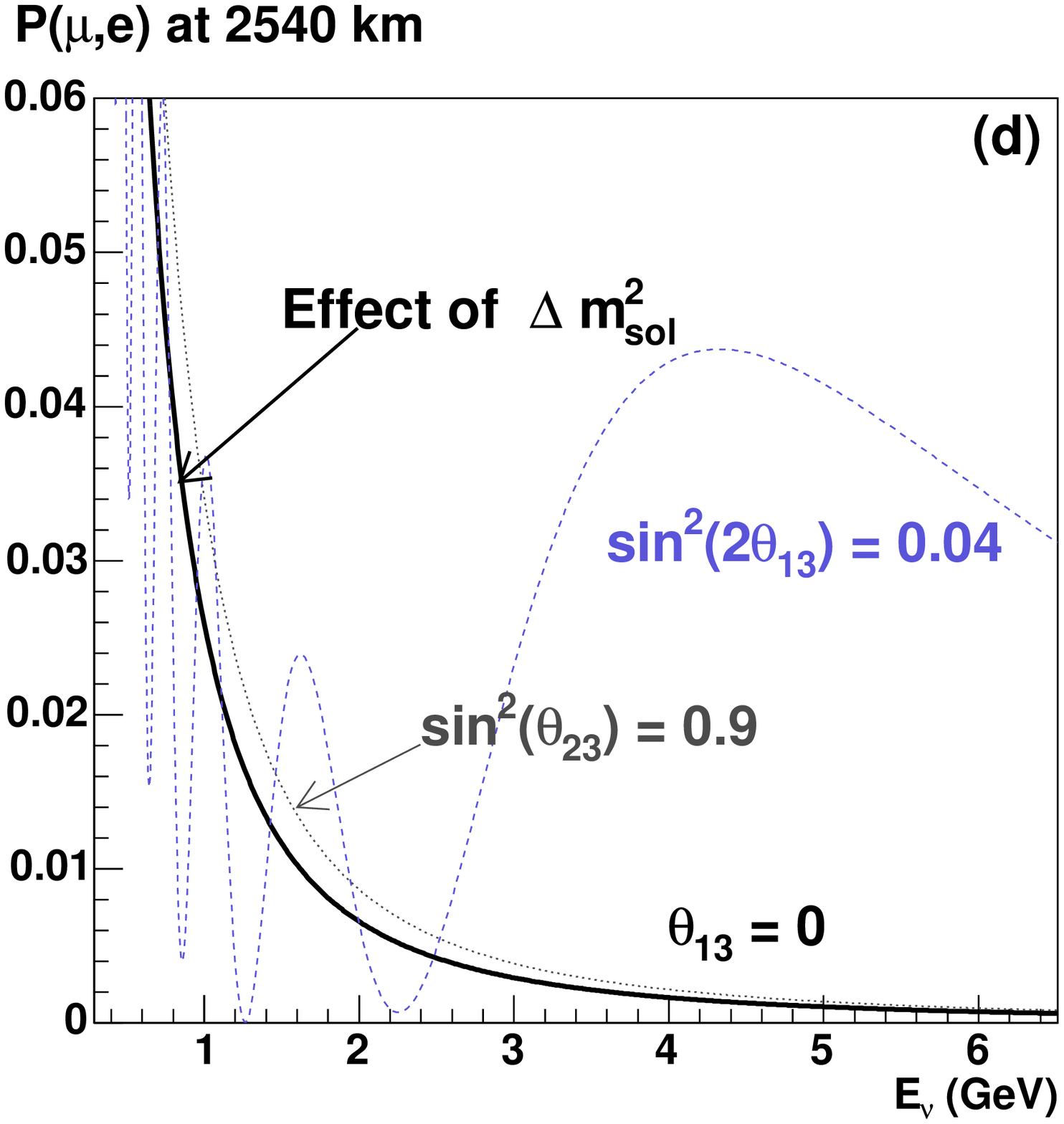}}{0.8}{0.8}
\end{minipage}
\vspace{-0.5cm}
\caption{Different effects contributing to the appearance probability over 
  the energy range covered by the BNL VLBNO.  
  (a) Current best fit for \dmatm{} and at $\pm$ 90\% CL (dot/dash resp.),
  (b) Oscillation in vacuum (solid line) and in  matter
  with \dmatm $>$ 0 (upper) and $<0$ (lower dashed),
  (c) the effect of varying CP phase angle and
  (d) the effect of $\theta_{13} = 0$.  See text for details.}
\label{prob}
\end{figure}

Figure \ref{prob} shows four affects to the \numutonue{} appearance
probability over the energy range covered by the VLBNO flux.  Figure
\ref{prob}.a shows that the features are well centered in the region of
flux coverage.  If the true value of \dmatm{} varies within current
uncertainties all features are still well within the flux coverage.

Figure \ref{prob}.b shows the large matter effect and that it appears
almost entirely in the first peak.  The effect due to the sign of
\dmatm{} will produce a clear result.

Figure \ref{prob}.c shows that the effect of a CP violating phase
increases as one goes to higher oscillations (lower energy).

Figure \ref{prob}.d shows that with $\theta_{13} = 0$ the BNL VLBNO
experiment is expected to see \numutonue{} appearance in the sub-GeV
region due to sub-dominant \dmsol{} driven oscillations.  This provides
a unique terrestrial based measurement of solar neutrino parameters
which can be checked against solar neutrino data from SK and SNO.  Any
deviation can be a sign of new physics.

Table \ref{table} summarizes these effects and illustrates in what
energy range they dominate.  The value of $\theta_{13}$ effects the
absolute event rate independently of energy.  The mass hierarchy
strongly influences the rates above 2 GeV.  The value of $\delta_{CP}$
is greatest in the middle energy range of 1-2 GeV.  Finally,
\numutonue{} appearance due to solar oscillations is very strong, but
only in the lower energy range below 1 GeV.  It is this distribution
of effects across the energy range that allows the BNL VLBNO
experiment to break the degeneracies.  Any LBNO experiment which
targets a single oscillation will not be able to disentangle these
effects without the proper addition of other detectors, baselines or
completely separate experiments.

\begin{figure}[htb]
  \begin{center}
\begin{minipage}{0.40\textwidth}
\overlay{{\tiny (a)}}{\includegraphics*[width=0.5\textwidth]{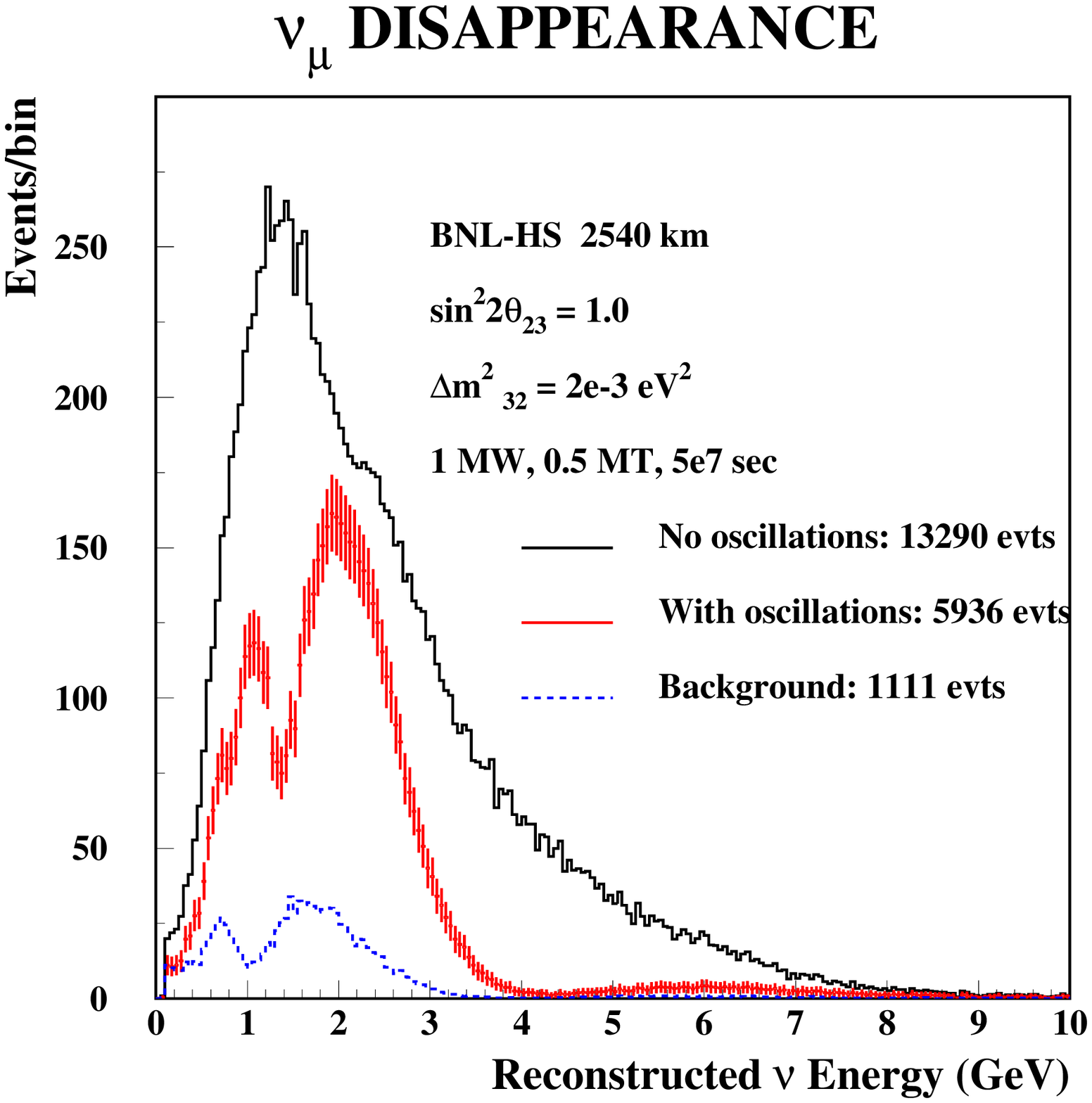}}{0.8}{0.8}%
\overlay{{\tiny (b)}}{\includegraphics*[width=0.5\textwidth]{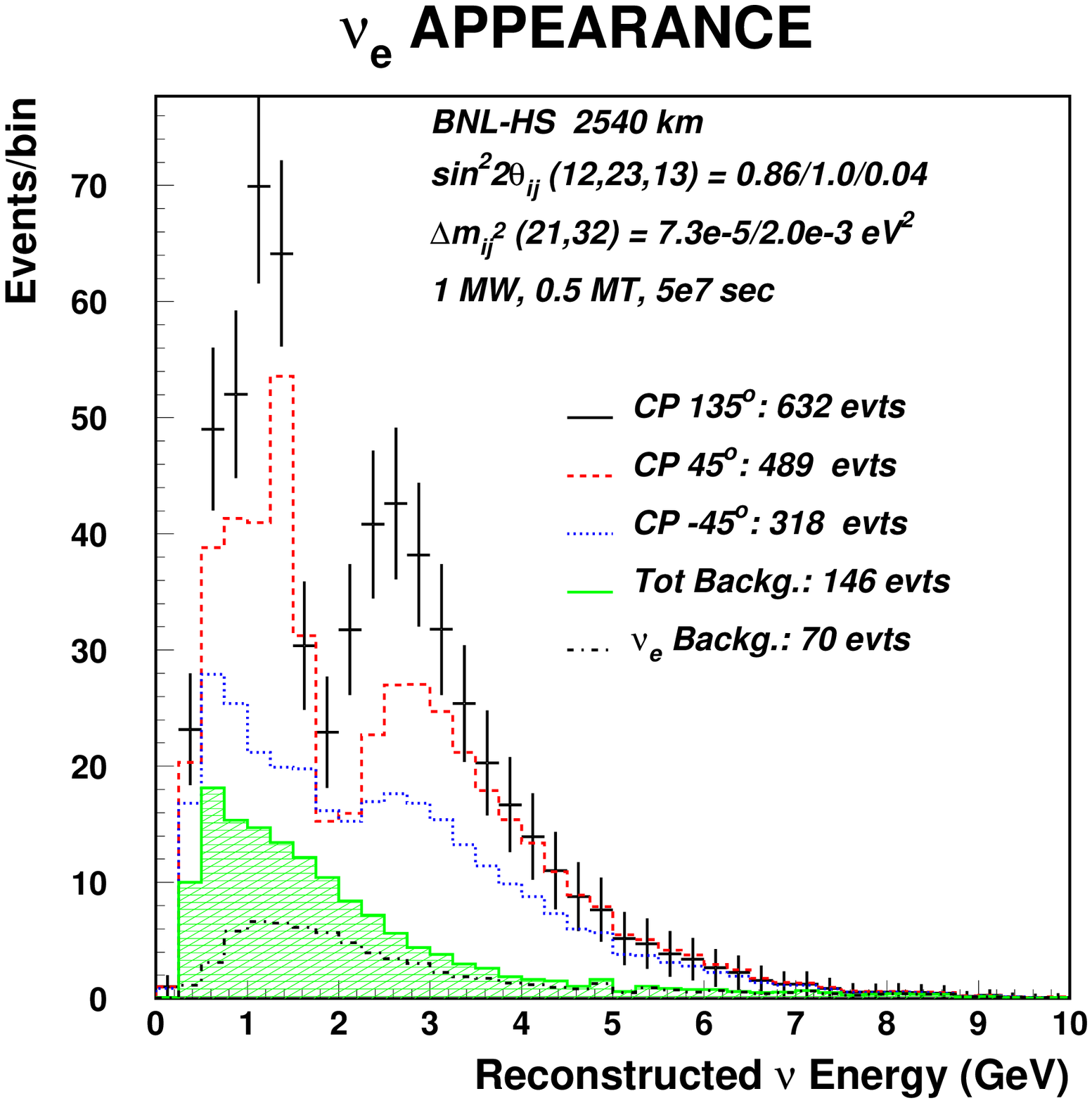}}{0.8}{0.8}

\overlay{{\tiny (c)}}{\includegraphics*[width=0.5\textwidth]{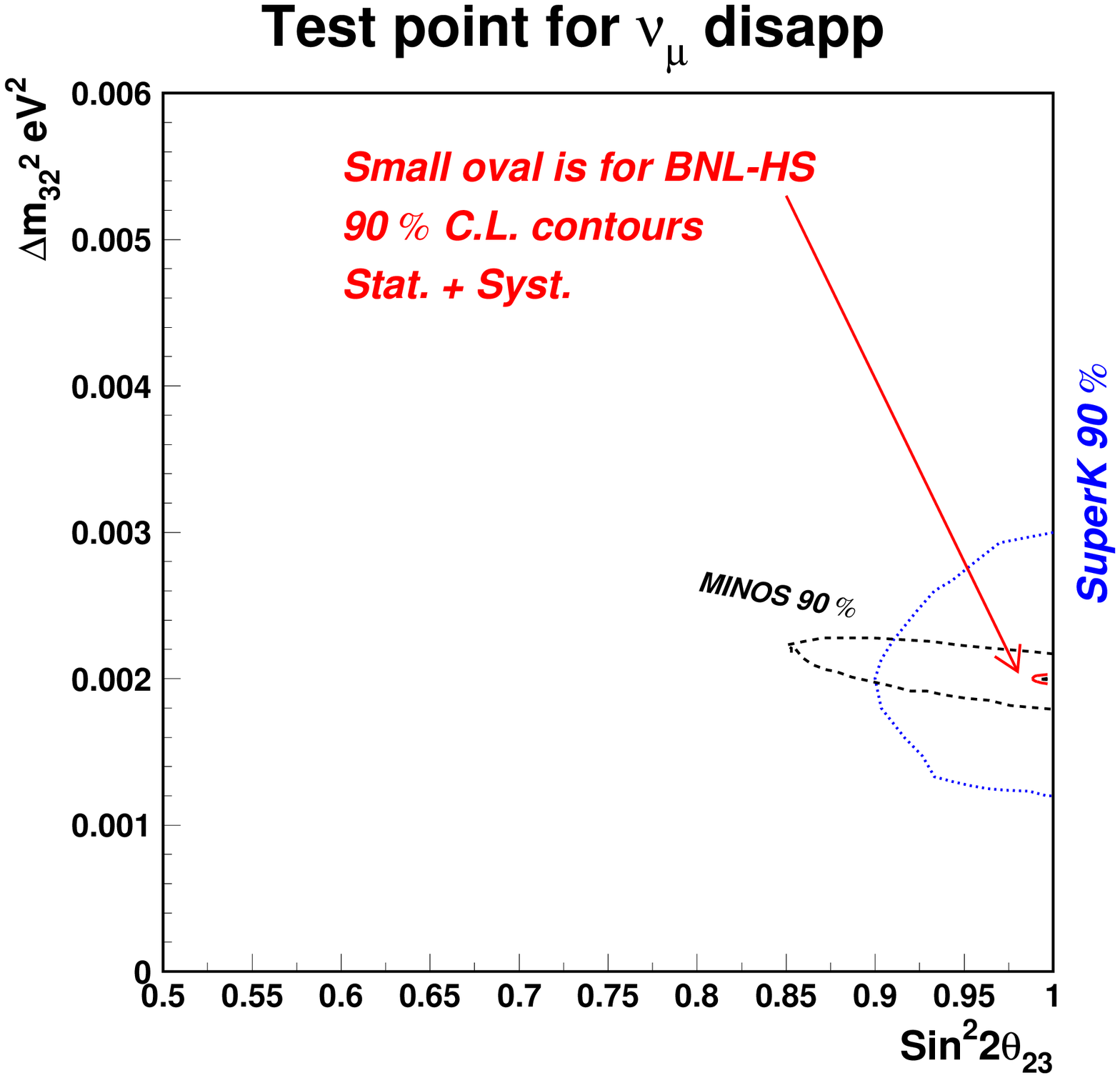}}{0.8}{0.8}%
\overlay{{\tiny (d)}}{\includegraphics*[width=0.5\textwidth]{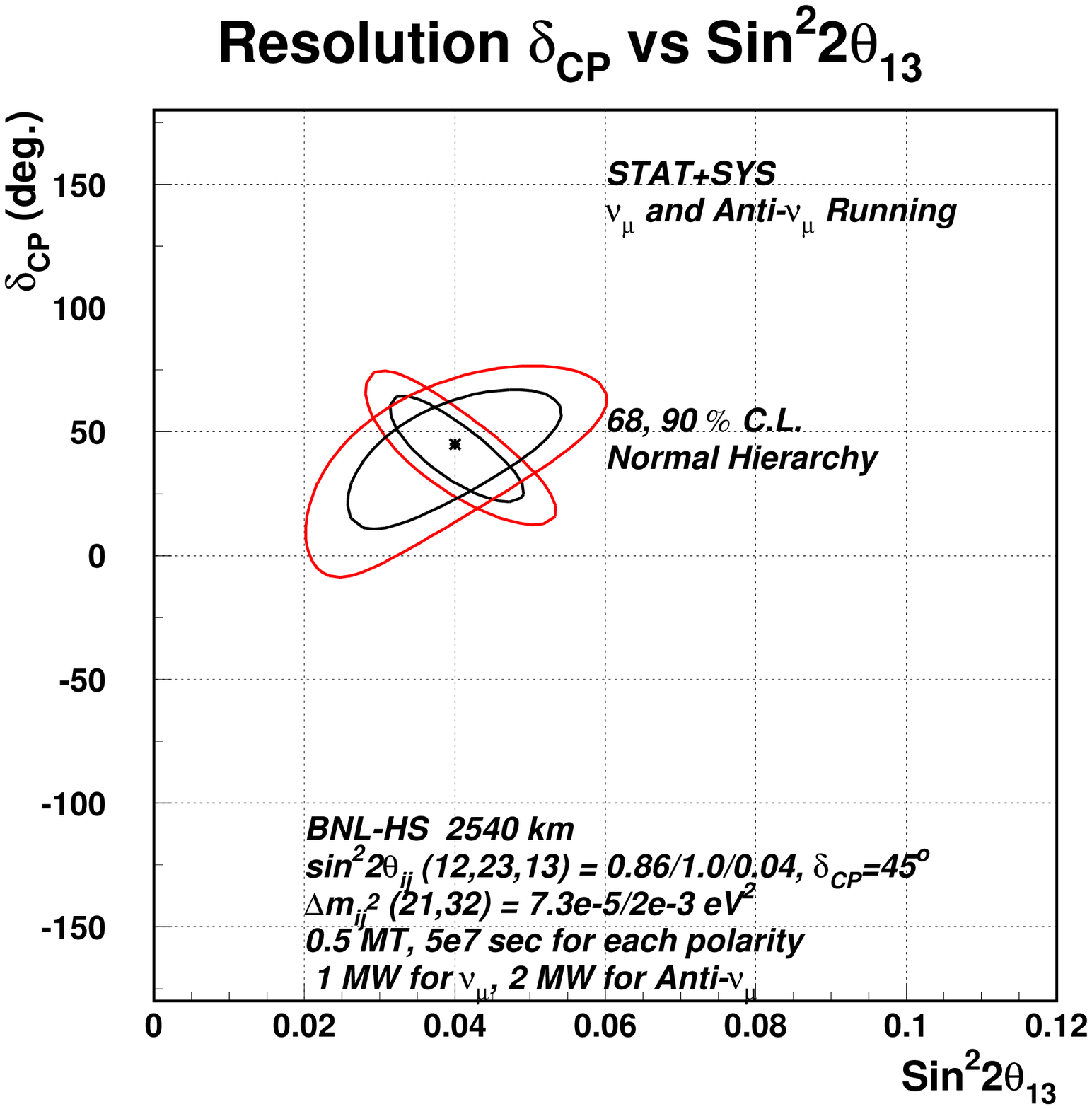}}{0.8}{0.8}

\end{minipage}
  \end{center}
\vspace{-1cm}
\caption{Expected disappearance (left) and appearance (right) spectra (top) and uncertainties for example test points (bottom).}
\label{res}
\end{figure}

The plots in Figure \ref{res} show expected spectra and parameter
resolutions.  The details of this study are presented
elsewhere\cite{milind}.

\section{Conclusion}

Through the use of a wide band and high energy neutrino beam, a very
long baseline and a massive far detector the BNL VLBNO experiment will
determine the mass hierarchy, precisely measure or strongly limit all
neutrino oscillation parameters and break parameter degeneracies.

\begin{table}[h]
\caption{Summary of strength of different appearance effects in different energy ranges.}
\begin{center}
    \begin{tabular}[ht]{|c|c|c|c|}
      \hline
      $E_\nu$ (GeV): & $<1$ & 1 - 2 & $>2$ \\
      \hline
      $\sin^22\theta_{13}$ & $\surd\surd$ & $\surd\surd$ & $\surd\surd$ \\
      \hline
      $sign(\Delta m^2_{32})$ & - & - & $\surd\surd\surd$ \\
      \hline
      $\delta_{CP}$ & $\surd$ & $\surd\surd$ & $\surd$ \\
      \hline
      solar & $\surd\surd\surd$ & $\surd$ & - \\
      \hline
    \end{tabular}
\label{table}
\end{center}
\end{table}

\begin{thebibliography}{9}


\bibitem{agsupgrade} W. T. Weng, M. Diwan, and D. Raparia (eds.), The
  AGS-Based Super Neutrino Beam CDR,
  BNL-73210-2004-IR, 2004. http://nwg.phy.bnl.gov/papers/agsnbcdr1.pdf
\bibitem{milind} M.V. Diwan, et al, VLBNO Experiment, Physics Review D 68, 012002 (2003).
\bibitem{chiaki} C. Yanagisawa, Water \cerenkov{} Simulation Studies on
  Backgrounds and Resolution, 3rd BNL/UCLA Workshop (2005).
\bibitem{prem} As taken from, I. Mocioiu, R. Shrock, Phys.Rev. D62
  (2000) 053017.
\end{thebibliography}
\end{document}